\documentclass[
    twocolumn,
	prd,
	amssymb,
	preprintnumbers,superscriptaddress,
	nofootinbib]{revtex4-1}

\pdfoutput=1
\usepackage{paralist}
\usepackage{graphicx}
\usepackage{enumitem}
\usepackage{latexsym}
\usepackage{amsfonts}
\usepackage{amssymb}
\usepackage{xcolor}
\usepackage[export]{adjustbox}
\usepackage{amsmath}
\usepackage[thinlines]{easytable}
\usepackage{slashed}
\usepackage{dcolumn}
\usepackage{verbatim}
\usepackage{paralist}
\usepackage{float}
\usepackage{multirow}
\usepackage{xspace}
\usepackage[normalem]{ulem}
\usepackage[
pdfauthor={Jeremy Sakstein}]{hyperref}
\usepackage{tabularx}
\usepackage{lettrine}
\usepackage{setspace}

\input Zallman.fd

\LettrineTextFont{\itshape}

\newcommand{\beq}{\begin{equation}}
\newcommand{\eeq}{\end{equation}}
\newcommand{\bea}{\begin{eqnarray}}
\newcommand{\eea}{\end{eqnarray}}

\newcommand{\refjnl}[1]{{\rm#1}}

\def\apj{\refjnl{ApJ}}                 
\def\aap{\refjnl{A\&A}}                

\def\mnras{\refjnl{MNRAS}}             
\usepackage{pdfbase}[2017/03/16]
\usepackage{xparse,ocgbase}
\usepackage{xcolor,calc}
\usepackage{tikzpagenodes,linegoal}
\usetikzlibrary{calc}
\usepackage{tcolorbox}

\ExplSyntaxOn
\let\tpPdfLink\pbs_pdflink:nn
\let\tpPdfAnnot\pbs_pdfannot:nnnn\let\tpPdfLastAnn\pbs_pdflastann:
\let\tpAppendToFields\pbs_appendtofields:n
\def\tpPdfXform{\pbs_pdfxform:nnnnn{1}{1}{}{}}
\let\tpPdfLastXform\pbs_pdflastxform:
\let\cListSet\clist_set:Nn\let\cListItem\clist_item:Nn
\ExplSyntaxOff

\usepackage{pdfbase}[2017/03/16]
\usepackage{xparse,ocgbase}
\usepackage{xcolor,calc}
\usepackage{tikzpagenodes,linegoal}
\usetikzlibrary{calc}
\usepackage{tcolorbox}

\ExplSyntaxOn
\let\tpPdfLink\pbs_pdflink:nn
\let\tpPdfAnnot\pbs_pdfannot:nnnn\let\tpPdfLastAnn\pbs_pdflastann:
\let\tpAppendToFields\pbs_appendtofields:n
\def\tpPdfXform{\pbs_pdfxform:nnnnn{1}{1}{}{}}
\let\tpPdfLastXform\pbs_pdflastxform:
\let\cListSet\clist_set:Nn\let\cListItem\clist_item:Nn
\ExplSyntaxOff

\makeatletter
\NewDocumentCommand{\tooltip}{%
  ssssO{\ifdefined\@linkcolor\@linkcolor\else blue\fi}mO{yellow!20}mO{0pt,0pt}%
}{{%
  \leavevmode%
  \IfBooleanT{#2}{%
    \ocgbase@new@ocg{tipOCG.\thetcnt}{%
      /Print<</PrintState/OFF>>/Export<</ExportState/OFF>>%
    }{false}%
    \xdef\tpTipOcg{\ocgbase@last@ocg}%
    \ocgbase@add@ocg@to@radiobtn@grp{tool@tips}{\ocgbase@last@ocg}%
  }%
  \tpPdfLink{%
    \IfBooleanTF{#4}{%
      /Subtype/Link/Border[0 0 0]/A <</S/SetOCGState/State [/Toggle \tpTipOcg]>>
    }{%
      /Subtype/Screen%
      /AA<<%
        \IfBooleanTF{#3}{%
          /E<</S/SetOCGState/State [/Toggle \tpTipOcg]>>%
        }{%
          \IfBooleanTF{#2}{%
            /E<</S/SetOCGState/State [/ON \tpTipOcg]>>%
            /X<</S/SetOCGState/State [/OFF \tpTipOcg]>>%
          }{
            \IfBooleanTF{#1}{%
              /E<</S/JavaScript/JS(%
                var fd=this.getField('tip.\thetcnt');%
                if(typeof(click\thetcnt)=='undefined'){%
                  var click\thetcnt=false;%
                  var fdor\thetcnt=fd.rect;var dragging\thetcnt=false;%
                }%
                if(fd.display==display.hidden){%
                  fd.delay=true;fd.display=display.visible;fd.delay=false;%
                }else{%
                  if(!click\thetcnt&&!dragging\thetcnt){fd.display=display.hidden;}%
                  if(!dragging\thetcnt){click\thetcnt=false;}%
                }%
                this.dirty=false;%
              )>>%
            }{%
              /E<</S/JavaScript/JS(%
                var fd=this.getField('tip.\thetcnt');%
                if(typeof(click\thetcnt)=='undefined'){%
                  var click\thetcnt=false;%
                  var fdor\thetcnt=fd.rect;var dragging\thetcnt=false;%
                }%
                if(fd.display==display.hidden){%
                  fd.delay=true;fd.display=display.visible;fd.delay=false;%
                }%
               this.dirty=false;%
              )>>%
              /X<</S/JavaScript/JS(%
                if(!click\thetcnt&&!dragging\thetcnt){fd.display=display.hidden;}%
                if(!dragging\thetcnt){click\thetcnt=false;}%
                this.dirty=false;%
              )>>%
            }%
            /U<</S/JavaScript/JS(click\thetcnt=true;this.dirty=false;)>>%
            /PC<</S/JavaScript/JS (%
              var fd=this.getField('tip.\thetcnt');%
              try{fd.rect=fdor\thetcnt;}catch(e){}%
              fd.display=display.hidden;this.dirty=false;%
            )>>%
            /PO<</S/JavaScript/JS(this.dirty=false;)>>%
          }%
        }%
      >>%
    }%
  }{{\color{#5}#6}}%
  \sbox\tiptext{%
    \IfBooleanT{#2}{%
      \ocgbase@oc@bdc{\tpTipOcg}\ocgbase@open@stack@push{\tpTipOcg}}%
    \tcbox[colframe=black,colback=#7,size=fbox,arc=1ex,sharp corners=southwest]{#8}%
    \IfBooleanT{#2}{\ocgbase@oc@emc\ocgbase@open@stack@pop\tpNull}%
  }%
  \cListSet\tpOffsets{#9}%
  \edef\twd{\the\wd\tiptext}%
  \edef\tht{\the\ht\tiptext}%
  \edef\tdp{\the\dp\tiptext}%
  \tipshift=0pt%
  \IfBooleanTF{#2}{%
    \setlength\whatsleft{\linegoal}%
  }{%
    \measureremainder{\whatsleft}%
  }%
  \ifdim\whatsleft<\dimexpr\twd+\cListItem\tpOffsets{1}\relax%
    \setlength\tipshift{\whatsleft-\twd-\cListItem\tpOffsets{1}}\fi%
  \IfBooleanF{#2}{\tpPdfXform{\tiptext}}%
  \raisebox{\heightof{#6}+\tdp+\cListItem\tpOffsets{2}}[0pt][0pt]{%
    \makebox[0pt][l]{\hspace{\dimexpr\tipshift+\cListItem\tpOffsets{1}\relax}%
    \IfBooleanTF{#2}{\usebox{\tiptext}}{%
      \tpPdfAnnot{\twd}{\tht}{\tdp}{%
        /Subtype/Widget/FT/Btn/T (tip.\thetcnt)%
        /AP<</N \tpPdfLastXform>>%
        /MK<</TP 1/I \tpPdfLastXform/IF<</S/A/FB true/A [0.0 0.0]>>>>%
        /Ff 65536/F 3%
        /AA <<%
          /U <<%
            /S/JavaScript/JS(%
              var fd=event.target;%
              var mX=this.mouseX;var mY=this.mouseY;%
              var drag=function(){%
                var nX=this.mouseX;var nY=this.mouseY;%
                var dX=nX-mX;var dY=nY-mY;%
                var fdr=fd.rect;%
                fdr[0]+=dX;fdr[1]+=dY;fdr[2]+=dX;fdr[3]+=dY;%
                fd.rect=fdr;mX=nX;mY=nY;%
              };%
              if(!dragging\thetcnt){%
                dragging\thetcnt=true;Int=app.setInterval("drag()",1);%
              }%
              else{app.clearInterval(Int);dragging\thetcnt=false;}%
              this.dirty=false;%
            )%
          >>%
        >>%
      }%
      \tpAppendToFields{\tpPdfLastAnn}%
    }%
  }}%
  \stepcounter{tcnt}%
}}
\makeatother
\newsavebox\tiptext\newcounter{tcnt}
\newlength{\whatsleft}\newlength{\tipshift}
\newcommand{\measureremainder}[1]{%
  \begin{tikzpicture}[overlay,remember picture]
    \path let \p0 = (0,0), \p1 = (current page.east) in
      [/utils/exec={\pgfmathsetlength#1{\x1-\x0}\global#1=#1}];
  \end{tikzpicture}%
}

\newcommand{\msun}{{\rm M}_\odot}
\newcommand{\rsun}{{\rm R}_\odot}

\DeclareRobustCommand{\okina}{%
  \raisebox{\dimexpr\fontcharht\font`A-\height}{%
    \scalebox{0.8}{`}%
  }%
}

\allowdisplaybreaks

\begin{document}

\title{Stars Would Remember a Recent Gravitational Transition:\\Testing G-step Resolutions of the Hubble Tension with Red Dwarfs}

\author{Jeremy Sakstein} \email{sakstein@hawaii.edu}
\affiliation{Department of Physics \& Astronomy, University of Hawai\okina i, Watanabe Hall, 2505 Correa Road, Honolulu, HI, 96822, USA}
\author{Harry Desmond}
\affiliation{Institute of Cosmology and Gravitation, University of Portsmouth, Dennis Sciama Building, Burnaby Road, Portsmouth PO1 3FX, UK\\}

\date{\today}

\begin{abstract}
We test the hypothesis that Newton's gravitational constant underwent a rapid transition approximately $73$--$130\,{\rm Myr}$ ago, as proposed in some resolutions of the Hubble tension.~Fully convective red dwarf stars with masses below $\sim0.35\,M_\odot$ have thermal timescales comparable to the proposed transition look-back times, so an appreciable present-day memory can remain.~We show that a rapid step in $G$ induces measurable changes in their radii and confront these predictions with mass--radius measurements of eight red dwarfs in detached eclipsing binaries.~We find that the $G$-step models proposed to resolve the Hubble tension, corresponding to a $4\%$ change in $G$ at a look-back time of $73\,{\rm Myr}$ and an $11\%$ change at $130\,{\rm Myr}$, are excluded at significances of $3.1\sigma$ and $4.0\sigma$, respectively.
\end{abstract}

\maketitle

\section{Introduction}

The Hubble tension remains one of the most pressing problems in cosmology \cite{Riess:2021jrx,Kamionkowski:2022pkx,Poulin:2023lkg,CosmoVerseNetwork:2025alb}.~Distance-ladder measurements of the Hubble constant $H_0$ are discrepant with inferences based on the cosmic microwave background (CMB) at the level of approximately $5\sigma$, depending on the tracers and data sets considered, with the CMB-inferred value being lower.~Possible explanations include the presence of unaccounted-for systematic effects \cite{Efstathiou:2020wxn,Mortsell:2021nzg,Mortsell:2021tcx,Riess:2021jrx,Sharon:2023ioz,Murakami:2023xuy,Riess:2023bfx,Bhardwaj:2023mau,Brout:2023wol,Dwomoh:2023bro,Uddin:2023iob,Riess:2024ohe,Freedman:2024eph,Riess:2024vfa}, or inadequacies in the statistical methodology~\citep{Feeney_2018_clarifying, Desmond_2026, Percival_2026, Stiskalek_2026_Cepheids, Stiskalek_2026_maser}.~However, the internal consistency of early-Universe measurements, together with a degree of concordance in distance-ladder estimates,
has made the possibility that the tension signals a failure of the standard cosmological model difficult to dismiss.

Proposed extensions of $\Lambda$CDM aimed at resolving the $H_0$ tension generally fall into two classes.~Early-universe resolutions invoke pre-recombination modifications to the Universe's evolution that raise the CMB-inferred value to match the distance-ladder measurement \cite{Poulin:2018cxd,Sakstein:2019fmf,Poulin:2023lkg,CosmoVerseNetwork:2025alb,Schoneberg:2026eys}.~Late-universe proposals instead invoke modifications at low redshift, including changes to the late-time expansion history \cite{Akarsu:2021fol,Raveri:2023zmr,Pedrotti:2025ccw,Bansal:2026axl} or to the calibration of the local distance ladder~\cite{Desmond:2019ygn,Desmond:2020wep,Marra:2021fvf,Perivolaropoulos:2021bds,Ruchika:2023ugh,Perivolaropoulos:2024yxv,Ruchika:2024ymt,Hogas:2023vim,Hogas:2023pjz,Hogas:2025mii}.

One mechanism for lowering the distance-ladder measurement is to introduce new physics that alters the properties of standard(izable) candles between rungs.~In models that resolve the Hubble tension, the resulting recalibration causes the conventional distance ladder to infer a value of $H_0$ that is systematically too high.~Correcting for this restores concordance with the CMB.~One example of this mechanism is screened modified gravity, which alters the calibration of Cepheids and tip of the red giant branch stars between rungs \cite{Desmond:2019ygn,Desmond:2020wep,Hogas:2023vim,Hogas:2023pjz,Hogas:2025mii}.~An alternative to screening is the $G$-step model (GSM) \cite{Marra:2021fvf,Perivolaropoulos:2021bds,Ruchika:2023ugh,Perivolaropoulos:2024yxv,Ruchika:2024ymt}, in which Newton's gravitational constant undergoes a rapid transition approximately $73$--$130$ Myr ago.\footnote{Strictly speaking, only dimensionless combinations of fundamental constants are directly observable.~The $G$-step should therefore be understood as a change in a dimensionless gravitational coupling, such as the gravitational fine-structure constant $\alpha_G=Gm_p^2/(\hbar c)$, under an assumed choice of fixed non-gravitational scales.~Following the convention of the existing literature, for simplicity we refer to this simply as a change in Newton's constant $G$.}~Such a
transition modifies the calibration of the local distance ladder.

In this paper, we test the $G$-step proposal using fully convective red dwarf stars with masses $M\lesssim0.35\,\msun$.~The characteristic thermal timescale on which these stars radiate away an amount of energy comparable to their gravitational binding energy is the Kelvin--Helmholtz timescale,
\begin{align}
\label{eq:tKH}
    t_{\rm KH}&\sim\frac{GM^2}{RL}
    \\&\simeq 3\,{\rm Gyr}\,
    \left(\frac{M}{0.1\,M_\odot}\right)^2
    \left(\frac{R}{0.1\,R_\odot}\right)^{-1}
    \left(\frac{L}{10^{-3}\,L_\odot}\right)^{-1},\nonumber
\end{align}
where the numerical estimate corresponds to a typical $0.1\,\msun$ red dwarf.~As shown below, the relaxation time of $G$-step induced radius perturbations is shorter, $t_{\rm relax}\simeq0.05\,t_{\rm KH}$, but remains comparable to the $73$--$130\,{\rm Myr}$ elapsed since the proposed transitions.~An appreciable present-day radius perturbation can therefore survive.

We find that, at fixed mass, a transition from a larger (smaller) value of $G$ to the present-day value leaves red dwarfs with radii that are inflated (deflated) relative to the equilibrium radii appropriate for the present-day value of $G$.~We compare numerical simulations of red dwarfs evolved under a $G$-step with mass--radius measurements of eight stars in detached eclipsing binaries.~We find that the $4\%$ and $11\%$ changes in $G$ proposed to resolve the Hubble tension are excluded at significances of $3.1\sigma$ and $4.0\sigma$, respectively.

This paper is organized as follows.~Section~\ref{sec:G-step-review} reviews the salient features of the $G$-step model and motivates the benchmark scenarios considered in this work.~In Section~\ref{sec:Analytic}, we develop an analytic model of red dwarfs evolving under a $G$-step and use it to build physical intuition for the effects of a rapid gravitational transition.~The observational data are described in Section~\ref{sec:data}, while the details of our numerical simulations are presented in Section~\ref{sec:MESA}.~Our statistical methodology is outlined in Section~\ref{sec:method}, the results are presented in Section~\ref{sec:results}, and our conclusions are given in Section~\ref{sec:concs}.

\section{Review of the $G$-Step Model}
\label{sec:G-step-review}

The $G$-step model aims to resolve the Hubble tension by inducing a systematic difference between the inferred intrinsic magnitudes of the calibrated and cosmological type-Ia supernovae (SNIa) samples.~Suppose that the intrinsic SNIa luminosity scales with Newton's constant as
\begin{equation}
    L\propto G^{\gamma}.
\end{equation}
If the calibrated SNIa sample lies after the transition, when $G=G_0$, while the cosmological sample lies before it, when $G=G_0(1+\Delta G/G_0)$, then the corresponding intrinsic luminosities, $L_{\rm cal}$ and $L_{\rm cos}$, satisfy
\begin{equation}
    \frac{L_{\rm cos}}{L_{\rm cal}}
    =
    \left(1+\frac{\Delta G}{G_0}\right)^\gamma.
\end{equation}
At fixed observed flux, the inferred luminosity distance scales as $d_L\propto L^{1/2}$, while at low redshift $H_0\propto d_L^{-1}$.~Analyzing the cosmological SNIa using the locally calibrated luminosity therefore gives
\begin{equation}
\label{deltaHGeneral}
    \frac{H_0^{\rm inf}}{H_0}
    =
    \left(\frac{L_{\rm cos}}{L_{\rm cal}}\right)^{1/2}
    =
    \left(1+\frac{\Delta G}{G_0}\right)^{\gamma/2}.
\end{equation}
The precise value of $\Delta G/G_0$ required to restore concordance therefore depends on the magnitude and sign of $\gamma$.~If the transition also affects the Cepheid calibrators, however, an additional correction arises because the Cepheid period--luminosity relation is itself sensitive to $G$ \cite{Sakstein:2019qgn,Ruchika:2023ugh}.

The original $G$-step model proposed a transition at $z=0.01$, corresponding to a look-back time of $130$ Myr \cite{Marra:2021fvf,Perivolaropoulos:2021bds}.~The transition lies between the cosmological and calibrated SNIa samples, so the Cepheid calibrators are unaffected.~Using the observed relation between SNIa peak luminosity and light-curve decay time, corresponding to an effective scaling $L_{\rm SNIa}\propto G^{1.46}$ \cite{Wright:2017rsu,Desmond:2019ygn}, the change required to restore concordance is $\Delta G/G_0\simeq+0.11$.

A more recent model \cite{Ruchika:2023ugh} instead proposes a transition at $22.4$ Mpc, corresponding to a look-back time of $73$ Myr.~This places the transition within the Cepheid-calibrated SNIa sample, so both the Cepheid calibration and the intrinsic SNIa luminosity are affected.~Using the scaling $L_{\rm Cep}\propto G^4$ for Cepheids \cite{Sakstein:2019qgn} together with $L_{\rm SNIa}\propto G^{1.46}$ for standardized SNIa \cite{Wright:2017rsu,Desmond:2019ygn}, Ref.~\cite{Ruchika:2023ugh} finds that restoring concordance requires $\Delta G/G_0\simeq+0.04$.

The GSM has motivated several studies of the phenomenological consequences of a rapid transition in $G$.~Ref.~\cite{Banik:2024yzi} argued that a transition occurring $130\,{\rm Myr}$ ago would modify the evolution of the Sun and Earth and lead to observable consequences for terrestrial and astrophysical quantities, including the climate and the year-to-day ratio.~These conclusions were subsequently disputed in Ref.~\cite{Perivolaropoulos:2025gzo}, which argued that several of the proposed constraints depend sensitively on the modeling of the solar and terrestrial response to the transition.~A separate challenge was raised in Ref.~\cite{Goldman:2024kot}, which argued that transitions with $\Delta G/G_0<0$ would cause neutron stars to contract to more compact configurations, releasing X-rays and neutrinos that are not observed.

A recent determination of the cosmological gravitational coupling using the cosmic microwave background, baryon acoustic oscillations, and big bang nucleosynthesis finds
$-0.025\leq\Delta G/G_0\leq0.048$ (95\% C.L.) \cite{Lamine:2024xno}.~If interpreted as a constraint on a universal Newton's constant, this disfavors some realizations of the GSM.~However, in theories with additional gravitational degrees of freedom, such as scalar--tensor theories, the coupling governing cosmological expansion need not coincide with the locally measured Newton's constant and may also depend on scale.~A fully consistent comparison therefore requires embedding the phenomenological $G$-step in an underlying theory.

Motivated by the considerations above, we consider two GSM benchmarks:
\begin{compactitem}
    \item[\textbf{Original GSM (OGSM):}] $\Delta G/G_0=+0.11$ at look-back time $t_{\rm step}=130$ Myr.
    \item[\textbf{Extended GSM (EGSM):}] $\Delta G/G_0=+0.04$ at look-back time $t_{\rm step}=73$ Myr.
\end{compactitem}

\section{Analytic Model of Red Dwarfs Evolving Under a $G$-step}
\label{sec:Analytic}

Before proceeding to detailed numerical modeling, we can gain intuition for how red dwarfs respond to a $G$-step by considering a simple analytic model.~Our goal is to determine how the stellar radius responds to a change in $G$ at fixed mass.~There are two effects that must be considered separately:~the equilibrium radius before and after the $G$-step, and the response of the star during the transition.~In both cases we use homology arguments in which radially dependent quantities are replaced by their characteristic scalings.

\subsection{Equilibrium Radius}

To begin, we take the hydrostatic equilibrium equation,
\begin{equation}
    \label{eq:HSE}
    \frac{\mathrm{d}P}{\mathrm{d}r}
    =-\frac{G M(r)\rho(r)}{r^2},
\end{equation}
and use $\rho_c\sim M/R^3$ to obtain the central-pressure scaling
\begin{equation}
\label{eq:Pc}
    P_c\sim \frac{GM^2}{R^4}.
\end{equation}
The red dwarfs studied in this work have $M\lesssim0.35\,\msun$ and are fully convective.~Approximating their interiors as an ideal gas,
\begin{equation}
\label{eq:ideal_gas_EOS}
   P= \frac{\rho k_BT}{\mu m_{u}},
\end{equation}
where $\mu$ is the mean molecular weight and $m_u$ is the atomic mass unit, gives
\begin{equation}
\label{eq:Tc}
    T_c \sim \frac{G M}{R}.
\end{equation}
The star is powered by pp-burning with energy generation rate per unit mass
\begin{equation}
    \label{eq:Lpp}
    \varepsilon_{\rm pp} \propto \rho T^4,
\end{equation}
so the luminosity from nuclear burning is 
\begin{equation}
    \label{eq:Lnuc}
    L_{\rm nuc}\sim M \varepsilon_{\rm pp} \sim G^4M^6R^{-7}.
\end{equation}
Stable equilibrium is achieved when the nuclear luminosity equals the surface luminosity, which is given by the Stefan--Boltzmann law
\begin{equation}
\label{eq:Lsurf}
    L_{\rm surf}=4\pi R^2\sigma T_{\rm eff}^4.
\end{equation}
Determining how $T_{\rm eff}$ scales with $G$, $M$, and $R$ requires matching the interior model to an atmospheric boundary condition, which yields (see Appendix \ref{sec:appendix})
\begin{equation}
    T_{\rm eff}^{49}
    \propto
    G^{13}M^7R.
    \label{eq:Teff49}
\end{equation}
Setting $L_{\rm nuc}=L_{\rm surf}$ with this relation gives
\begin{equation}
    \label{eq:RvsG}
    R\propto G^\frac{144}{445}M^\frac{266}{445}\simeq G^{0.324}M^{0.598}.
\end{equation}
Thus, after a $G$-step from a larger (smaller) $G$ to today's value the new equilibrium radius is smaller (larger).

\subsection{Response to the $G$-step}

The equilibrium scaling in Eq.~\eqref{eq:RvsG} does not describe the rapid response of the star to the $G$-step.~Because the transition and subsequent dynamical readjustment occur on a timescale much shorter than the Kelvin--Helmholtz timescale, the star has insufficient time to thermally relax and its entropy profile changes little.~The response can therefore be approximated as adiabatic.

A fully convective star can be approximated as an $n=3/2$ polytrope \cite{2012sse..book.....K,Sakstein:2015zoa,Sakstein:2015aac},
\begin{equation}
    P=K\rho^{5/3},
\end{equation}
where $K$ is related to the specific entropy.~The Lane--Emden relations
give
\begin{equation}
    K\propto G R M^{1/3}.
\end{equation}
Since $K$ changes little during the rapid transition, at fixed mass we
therefore have
\begin{equation}
    R\propto G^{-1}.
    \label{eq:Radiabatic}
\end{equation}
Thus, we expect a rapid expansion (contraction) as $G$ transitions from a larger (smaller) value to its present-day value.

\subsection{Expected Evolution}

Together, the arguments above imply the following evolution for a red dwarf experiencing a $G$-step.~Before the transition, a star with a larger (smaller) value of $G$ has a larger (smaller) equilibrium radius than a star with the same mass evolving under the present-day value of $G$.~As $G$ transitions to its present-day value, the star undergoes a rapid quasi-adiabatic expansion (contraction).~It subsequently relaxes toward the new equilibrium radius through thermal readjustment.

The relaxation time of a small radius perturbation can be estimated from the homology relations above.~At fixed $G$ and $M$, Eqs.~\eqref{eq:Lnuc} and \eqref{eq:Teff49} imply
\begin{equation}
    L_{\rm nuc}\propto R^{-7},
    \qquad
    L_{\rm surf}\propto R^{102/49},
\end{equation}
where the second relation follows from $L_{\rm surf}\propto R^2T_{\rm eff}^4$ and $T_{\rm eff}\propto R^{1/49}$.~Writing $R=R_{\rm eq}(1+\delta)$ and linearizing about equilibrium, where $L_{\rm nuc}=L_{\rm surf}\equiv L$, therefore gives
\begin{equation}
    L_{\rm nuc}-L_{\rm surf}
    \simeq
    -\left(7+\frac{102}{49}\right)L\delta.
\end{equation}
For an $n=3/2$ polytrope the total stellar energy is \cite{2009itss.book.....P,2012sse..book.....K}
\begin{equation}
    E=-\frac{3}{7}\frac{GM^2}{R},
\end{equation}
so, to linear order,
\begin{equation}
    \frac{dE}{dt}
    \simeq
    \frac{3}{7}\frac{GM^2}{R}\dot{\delta}.
\end{equation}
Using $dE/dt=L_{\rm nuc}-L_{\rm surf}$ together with the definition of $t_{\rm KH}$ in Eq.~\eqref{eq:tKH}, we obtain
\begin{equation}
    \dot{\delta}
    =
    -\frac{7}{3}
    \left(7+\frac{102}{49}\right)
    \frac{\delta}{t_{\rm KH}},
\end{equation}
and hence
\begin{equation}
    t_{\rm relax}
    \sim
    \frac{3}{7}
    \left(7+\frac{102}{49}\right)^{-1}
    t_{\rm KH}
    \simeq 0.05\,t_{\rm KH}.
\end{equation}
For the low-mass stars considered here this corresponds to a characteristic relaxation time of order $10^8\,{\rm yr}$, comparable to the $73$--$130\,{\rm Myr}$ elapsed since the proposed transitions.~We therefore expect an appreciable fraction of the radius perturbation to remain at the present day, as borne out by our numerical simulations.

\section{Data}
\label{sec:data}

We compare our stellar evolution models with detached eclipsing binaries drawn from Ref.~\cite{2024MNRAS.528.5703S}.~Detached eclipsing binaries provide some of the most precise measurements of stellar masses and radii, making them ideal laboratories for testing stellar evolution models and new physics \cite{Zentner:2011wx,Sakstein:2015aac,Sakstein:2015zoa,Croon:2024waz,2026arXiv260602732D}.~To facilitate comparison with models of isolated fully convective stars, we restrict the sample to systems with $0.1 \leq M/\msun \leq 0.31$, orbital periods $P\geq5$ days, and available mass--radius covariance information.~The mass cut selects stars expected to be fully convective in standard stellar evolution models \cite{Chabrier:1997vx,2015A&A...577A..42B,2018A&A...619A.177B}, while the period cut reduces the effects of tidal locking and irradiation from the primary star.~These requirements leave a sample of eight stars.~The stellar masses, radii, metallicities, and associated uncertainties are taken from Ref.~\cite{2024MNRAS.528.5703S}, and are given in Appendix~\ref{app:data}.~The systems are observed after the gravitational coupling has returned to its present-day value $G_0$, so the orbital dynamics at the epoch of observation are governed by $G_0$.~We therefore use the published stellar parameters directly in our analysis.

\begin{figure*}[ht]
    \centering
    \includegraphics[width=0.85\textwidth]{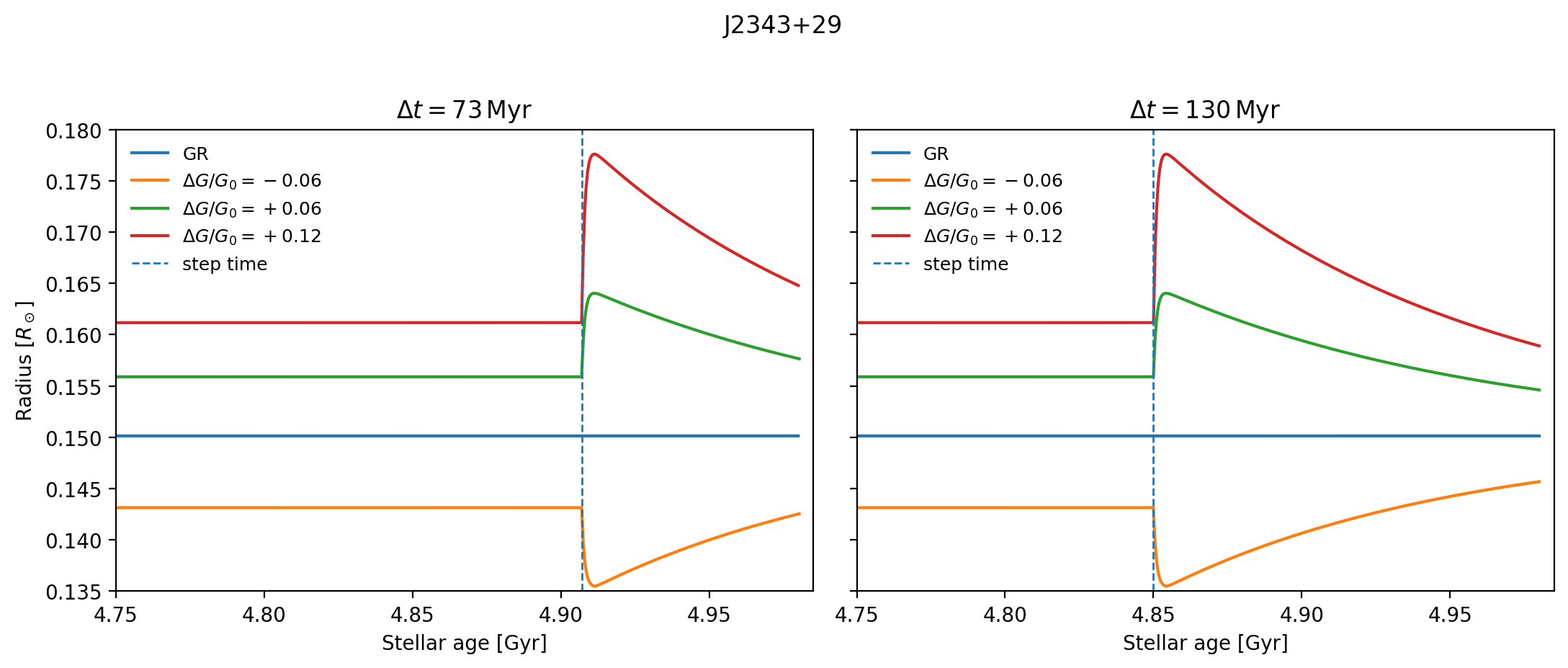}
    \caption{Radius vs.~time for J2343+29 for representative positive and negative changes in Newton's constant given in the figure and $f_{\rm spot}=0.34$.~The left (right) panel shows the EGSM (OGSM) benchmark with a $G$-step 73 Myr (130 Myr) ago.~The vertical dashed blue line shows the time of the $G$-step.}
    \label{fig:mesa}
\end{figure*}

\section{Stellar Modeling}
\label{sec:MESA}

Our simulations are performed using the stellar evolution code \textsc{MESA}, version 23.05.1 \cite{Paxton:2010ji,Paxton:2013pj,Paxton:2015jva,Paxton:2017eie,Paxton:2019lxx,MESA:2022zpy}, with a custom implementation of a $G$-step at an arbitrary stellar age.~The inlists and run\_star\_extras needed to reproduce our results are included in the paper's reproduction package \cite{sakstein_2026_23093223}.

The $G$-step is implemented by evolving each star from the pre-main sequence using
\begin{equation}
    G(t)=G_0\left[
    1+\frac{\Delta G}{G_0}
    \left(1-e^{-t/\tau_s}\right)
    \right],
\end{equation}
where $\tau_s=1\,{\rm Myr}$.~This initial ramp from $G_0$ to $G_0+\Delta G$ avoids numerical convergence problems.~During the first five e-foldings, the maximum timestep is limited to $0.1\,\tau_s$.

We evolve all models to a common final age of $4.98\,{\rm Gyr}$ and place the transition at
\begin{equation}
    t_{\rm step}=4.98\,{\rm Gyr}-\Delta t,
\end{equation}
where $\Delta t=73$ or $130\,{\rm Myr}$ for the two benchmarks.~At the transition, $G$ relaxes to its present value according to
\begin{equation}
    G(t)=G_0\left[
    1+\frac{\Delta G}{G_0}
    e^{-(t-t_{\rm step})/\tau_{\rm step}}
    \right],
    \qquad t\geq t_{\rm step},
\end{equation}
with $\tau_{\rm step}=1\,{\rm Myr}$.~We tested the sensitivity of the present-day stellar radius to this smoothing prescription by varying $\tau_{\rm step}$ over $0.1\,{\rm Myr}\leq\tau_{\rm step}\leq2\,{\rm Myr}$ and found that the final radius changes by less than $0.04\%$.~Our results are therefore insensitive to the precise value of $\tau_{\rm step}$.~

To resolve the transition, the maximum timestep is limited to $1\,{\rm Myr}$ during the $50\,{\rm Myr}$ preceding $t_{\rm step}$ and to $2\times10^4\,{\rm yr}$ during its first five e-foldings.~The integration is forced to land exactly at $t_{\rm step}$ and at the final age, preventing differences between models arising from the placement of nonuniform timesteps.~Since fully convective red dwarfs evolve only weakly after reaching the main sequence \cite{2015A&A...577A..42B}, the absolute choice of the common final age is unimportant provided that $\Delta t$ is held fixed.

\subsection{Magnetic Activity and Starspots}

Magnetic activity can modify the radii of low-mass stars through the presence of cool starspots.~Starspots are regions of the photosphere in which strong magnetic fields suppress convective energy transport, reducing the local temperature and emergent flux relative to the surrounding photosphere.~Blocking a fraction of the surface flux inhibits the star's ability to radiate its interior luminosity, which causes the stellar structure to adjust to a larger radius.~This mechanism has been shown to produce significant radius inflation in magnetically active low-mass stars and eclipsing binaries \cite{Chabrier:2007ue,2015ApJ...807..174S,MESA:2022zpy}.

Starspots are particularly important for our analysis because their effect is partially degenerate with that of a $G$-step.~Changing $G$ alters the mass--radius relation, while increasing the spot covering fraction inflates the stellar radius at fixed mass.~

We model starspots using the \textsc{MESA} two-parameter prescription \cite{MESA:2022zpy} introduced by \cite{2015ApJ...807..174S}.~In this description, the fraction of the stellar surface covered by spots is given by $f_{\rm spot}\in[0,1]$ and the ratio of the spot temperature to that of the unspotted photosphere is specified by $x_{\rm spot}=T_{\rm spot}/T_{\rm amb}$, where $T_{\rm spot}$ is the spot temperature and $T_{\rm amb}$ is the temperature of the unspotted photosphere.~We fix $x_{\rm spot}=0.85$ \cite{MESA:2022zpy} and allow $f_{\rm spot}$ to vary independently for each star.~For this choice of temperature contrast, the fraction of the photospheric flux blocked by spots is approximately
\begin{equation}
    \beta=f_{\rm spot}(1-x_{\rm spot}^4)\simeq0.48f_{\rm spot}.
\end{equation}

\subsection{Other Stellar Physics}

For fully convective stars in the mass range considered here, variations in individual stellar-model ingredients, including the equation of state, mixing length, radiative opacity, atmospheric boundary condition, and initial helium abundance, generally produce percent-level or smaller changes in the predicted radii \cite{2018MNRAS.476...27T}.~We therefore adopt fiducial choices for these ingredients.~Specifically, we use the default \textsc{MESA} equation-of-state blend and opacity tables \cite{Paxton:2010ji,MESA:2022zpy}, with a mixing-length parameter $\alpha_{\rm MLT}=2.0$.~For the outer boundary condition we use the \texttt{tau\_100} model-atmosphere tables, appropriate for the low-mass objects we study \cite{Paxton:2010ji}.~The initial helium abundance is set using the \textsc{MESA} relation $Y=0.24+2Z$.~We tested representative variations in the atmosphere, mixing length, and equation of state and similarly found percent-level or smaller changes in the predicted radii, consistent with Ref.~\cite{2018MNRAS.476...27T}.

\section{Method}
\label{sec:method}

We compare the observed masses and radii with the predictions of our \textsc{MESA} models using a profile-likelihood analysis.~The analysis code required to reproduce the profile-likelihood and bootstrap calculations is included in the reproduction package \cite{sakstein_2026_23093223}.

For each star $i$, we account for the correlation between the measured mass and radius by defining
\begin{align}
    x_i&=\frac{m_i-M_i}{\sigma_{M_i}},
    \\
    y_i&=\frac{R_{\rm th}(m_i,Z_i,\Delta G/G_0,f_{{\rm spot},i})-R_i}{\sigma_{R_i}},
\end{align}
where $M_i$ and $R_i$ are the measured mass and radius, $\sigma_{M_i}$ and $\sigma_{R_i}$ are their marginal uncertainties, and $m_i$ is the unknown true stellar mass.~The theoretical radius $R_{\rm th}$ is obtained from the \textsc{MESA} grid at the metallicity, $G$-step parameters, and spot covering fraction appropriate to each model.~The contribution of each star to the likelihood is then
\begin{equation}
\label{eq:chi2_cov}
    \chi_i^2
    =
    \frac{x_i^2+y_i^2-2\rho_i x_i y_i}
         {1-\rho_i^2},
\end{equation}
where $\rho_i$ is the measured mass--radius correlation coefficient.

For each system, we determine the mass--radius correlation coefficient $\rho_i$ from the posterior samples provided by Ref.~\cite{2024MNRAS.528.5703S}.\footnote{We are grateful to M.~Swayne for providing the posterior samples used to determine the mass--radius correlations.}~Together with the published marginal uncertainties $\sigma_{M_i}$ and $\sigma_{R_i}$, this specifies the mass--radius covariance matrix for each system.~We verified that the joint mass--radius posterior samples are approximately Gaussian, so the covariance matrix provides an adequate description for the likelihood analysis.~Accounting for this covariance is particularly important because changing $G$ alters the predicted mass--radius relation, whereas changing $f_{\rm spot}$ primarily shifts the radius at fixed mass.~Since the measured masses and radii are strongly correlated for several systems, the likelihood of these different shifts depends sensitively on their direction in the mass--radius plane.~The correlations are all positive and range from $\rho_i=0.53$ to $0.90$, with three systems, J1928-38, J2046+06, and J2343+29, having $\rho_i>0.85$.

For each target, we calculate models at a set of masses spanning the observationally allowed range and linearly interpolate the predicted radius between adjacent mass nodes.~At fixed $\Delta G/G_0$ and $f_{{\rm spot},i}$, we profile over the unknown true mass $m_i$ by minimizing Eq.~\eqref{eq:chi2_cov} analytically along each piecewise-linear segment of the interpolated mass--radius relation and retaining the global minimum.~The metallicity of each stellar model is fixed to the central value listed in Table~\ref{tab:sample}.~As a sensitivity test, varying the metallicity of each system by its observational $1\sigma$ uncertainty changes the predicted radius by at most $0.26\%$, corresponding to at most $0.17\sigma_R$, and therefore has a negligible effect on the likelihood.~

At each value of $\Delta G/G_0$, the spot covering fraction of each star is treated as an independent nuisance parameter constrained to its full physical range, $0\leq f_{{\rm spot},i}\leq1$.~The profile chi-square is therefore
\begin{equation}
    \chi^2_{\rm prof}\left(\frac{\Delta G}{G_0}\right)
    =
    \sum_i
    \min_{m_i,\,f_{{\rm spot},i}}
    \chi_i^2
    \left(
        \frac{\Delta G}{G_0},
        f_{{\rm spot},i},
        m_i
    \right).
\end{equation}
We evaluate this quantity over a grid of positive and negative values of $\Delta G/G_0$ separately for the two transition times, $\Delta t=73$ and $130\,{\rm Myr}$, with standard gravity corresponding to $\Delta G/G_0=0$.

The role of the mass--radius covariance can be seen particularly clearly for systems whose preferred spot covering fraction reaches the boundary $f_{{\rm spot},i}=0$.~Along a linear segment of the theoretical mass--radius relation, write
\begin{equation}
    y_i=y_{0,i}+s_i x_i,
\end{equation}
where
\begin{equation}
    y_{0,i}
    =
    \frac{R_{\rm th}(M_i)-R_i}{\sigma_{R_i}},
    \qquad
    s_i
    =
    \frac{dR_{\rm th}}{dm}
    \frac{\sigma_{M_i}}{\sigma_{R_i}}.
\end{equation}
Minimizing Eq.~\eqref{eq:chi2_cov} over the stellar mass gives
\begin{equation}
    \chi^2_{i,\min}
    =
    \frac{y_{0,i}^2}
    {1+s_i^2-2\rho_i s_i},
\end{equation}
provided that the minimum lies within the interpolated mass segment.~Thus the mass--radius covariance directly affects the penalty associated with an irreducible radius residual and can strengthen the constraint when the model displacement lies along the narrow direction of the observational error ellipse.

We define the likelihood-ratio statistic
\begin{equation}
    q\left(\frac{\Delta G}{G_0}\right)
    =
    \chi^2_{\rm prof}\left(\frac{\Delta G}{G_0}\right)
    -
    \chi^2_{\rm min},
\end{equation}
where $\chi^2_{\rm min}$ is the minimum obtained when $\Delta G/G_0$ is also allowed to vary.~For both transition times considered here, the global minimum occurs at $\Delta G/G_0=0$.~Because the nuisance parameters $f_{{\rm spot},i}$ are bounded below by zero and several of them reach this boundary, the regularity conditions underlying Wilks' theorem are not satisfied.~We therefore calibrate the significance of each benchmark using a parametric bootstrap.~For a fixed benchmark value of $\Delta G/G_0$, we first determine the conditional maximum-likelihood values of the stellar masses and spot covering fractions.~We then generate synthetic mass--radius measurements from the corresponding correlated Gaussian likelihood and refit each realization both with $\Delta G/G_0$ fixed to the benchmark value and with $\Delta G/G_0$ allowed to vary.~The fraction of realizations for which the simulated likelihood-ratio statistic exceeds the observed value gives the $p$-value for that benchmark.

\begin{figure}[t]
    \centering
    \includegraphics[width=0.48\textwidth]{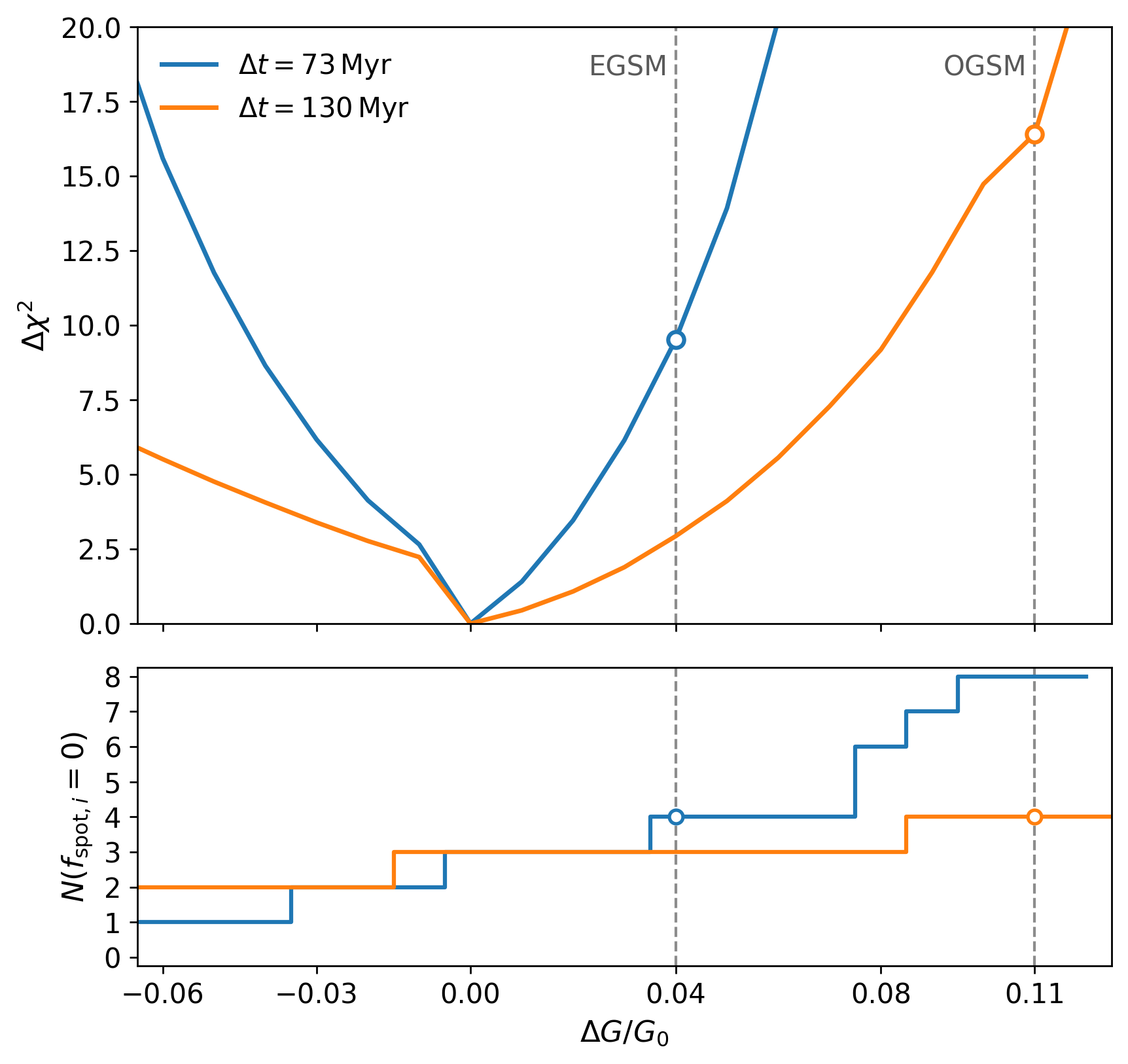}
    \caption{\textbf{Upper panel:} Profile likelihood $\Delta\chi^2$ as a function of $\Delta G/G_0$ for transition look-back times of $73\,{\rm Myr}$ and $130\,{\rm Myr}$.~The dashed vertical lines mark the EGSM and OGSM benchmark amplitudes, $\Delta G/G_0=0.04$ and $0.11$, respectively, and the circles indicate the corresponding values of $\Delta\chi^2$.~\textbf{Lower panel:} Number of systems for which the independently profiled spot covering fraction reaches the physical boundary $f_{{\rm spot},i}=0$.~The stellar mass and spot covering fraction are profiled independently for each system at each value of $\Delta G/G_0$.}
    \label{fig:likelihood}
\end{figure}

\section{Results}
\label{sec:results}

Figure~\ref{fig:mesa} shows representative results from our numerical simulations for J2343+29 with $f_{\rm spot}=0.34$.~The figure follows the stellar radius from before the transition to the present day for transition times corresponding to the EGSM and OGSM benchmarks and for representative positive and negative values of $\Delta G/G_0$.~The qualitative features predicted analytically in Section~\ref{sec:Analytic}---the change in the equilibrium radius, the rapid quasi-adiabatic response to the $G$-step, and the subsequent thermal relaxation---are reproduced in the numerical evolution.

The results of the profile-likelihood analysis are shown in the upper panel of Figure~\ref{fig:likelihood}.~For both transition times, the profile likelihood is minimized at $\Delta G/G_0=0$, with no preference for a nonzero recent change in $G$.~At the EGSM benchmark, $\Delta G/G_0=0.04$ for a look-back time of $73\,{\rm Myr}$, we find $\Delta\chi^2=9.52$.~At the OGSM benchmark, $\Delta G/G_0=0.11$ for a look-back time of $130\,{\rm Myr}$, we find $\Delta\chi^2=16.41$.

As a conservative test of residual stellar-model systematics, we also considered a coherent $1\%$ downward shift in the predicted radii, which acts in the direction most degenerate with a positive $G$-step.~The likelihood-ratio statistics at the EGSM and OGSM benchmarks are then reduced from $\Delta\chi^2=9.52$ and $\Delta\chi^2=16.41$ to $\Delta\chi^2=5.72$ and $\Delta\chi^2=10.92$, respectively.~Thus, stellar-model systematics at the level expected for these stars can weaken the statistical significance of the constraints, but our tests provide no indication that they would overturn the qualitative conclusion that the proposed $G$-step benchmarks are disfavored.

The behavior of the profile likelihood at positive $\Delta G/G_0$ is closely connected to the starspot covering fractions.~Because each system has its own independently profiled spot covering fraction, systems whose measured mass and radius can be reproduced with an interior value $0<f_{{\rm spot},i}<1$ contribute negligibly to the profile chi-square.~The sensitivity to positive $\Delta G/G_0$ therefore arises when the preferred spot covering fraction reaches the physical boundary $f_{{\rm spot},i}=0$.~At both benchmark amplitudes, four of the eight systems lie at this boundary.~Once a system reaches this boundary, the radius shift produced by increasing $\Delta G/G_0$ can no longer be compensated by further reducing its spot coverage, and its contribution is governed by the correlated mass--radius residual described in Section~\ref{sec:method}.~At this boundary the model is unspotted and hence independent of the assumed spot temperature contrast:~$\beta=f_{\rm spot}(1-x_{\rm spot}^4)=0$ for any $x_{\rm spot}$, so varying $x_{\rm spot}$ cannot provide the further reduction in spot-induced radius inflation that would be required to compensate a positive $G$-step.

We find $47$ of $50{,}000$ realizations with a likelihood-ratio statistic at least as large as the observed value for the EGSM benchmark, corresponding to $p=(9.4\pm1.4)\times10^{-4}$, or a one-sided Gaussian significance of approximately $3.1\sigma$.~For the OGSM benchmark, $18$ of $550{,}000$ realizations exceed the observed value, corresponding to $p=(3.3\pm0.8)\times10^{-5}$, or approximately $4.0\sigma$.~The quoted uncertainties reflect the Monte Carlo counting uncertainty of the bootstrap calibration.~The two Hubble-tension-motivated benchmarks are therefore excluded at approximately $3.1\sigma$ and $4.0\sigma$, respectively.

As a robustness check, we repeated the analysis after removing each system in turn. The profile likelihood remains minimized at $\Delta G/G_0=0$ in every case. Omitting J2343+29 produces the largest reduction in the benchmark likelihood ratio, from $q=9.52$ to $4.91$ for the EGSM benchmark and from $q=16.41$ to $3.96$ for the OGSM benchmark. By comparison, omitting any other individual system leaves $q\geq5.67$ for the EGSM benchmark and $q\geq13.53$ for the OGSM benchmark. Repeating the parametric-bootstrap calibration for the seven-star sample without J2343+29 gives $p=7.14\times10^{-3}$ ($2.45\sigma$) for EGSM and $p=2.87\times10^{-2}$ ($1.90\sigma$) for OGSM. Conservatively accounting for realizations that reach the finite stellar-mass grid boundaries changes these to $p=(7.04$--$7.96)\times10^{-3}$ and $p=(2.832$--$3.234)\times10^{-2}$, respectively. Thus J2343+29 carries substantial statistical weight, particularly for the OGSM constraint, although no single-star omission shifts the preferred value away from $\Delta G/G_0=0$.

\begin{figure}[t]
    \centering
\includegraphics[width=0.48\textwidth]{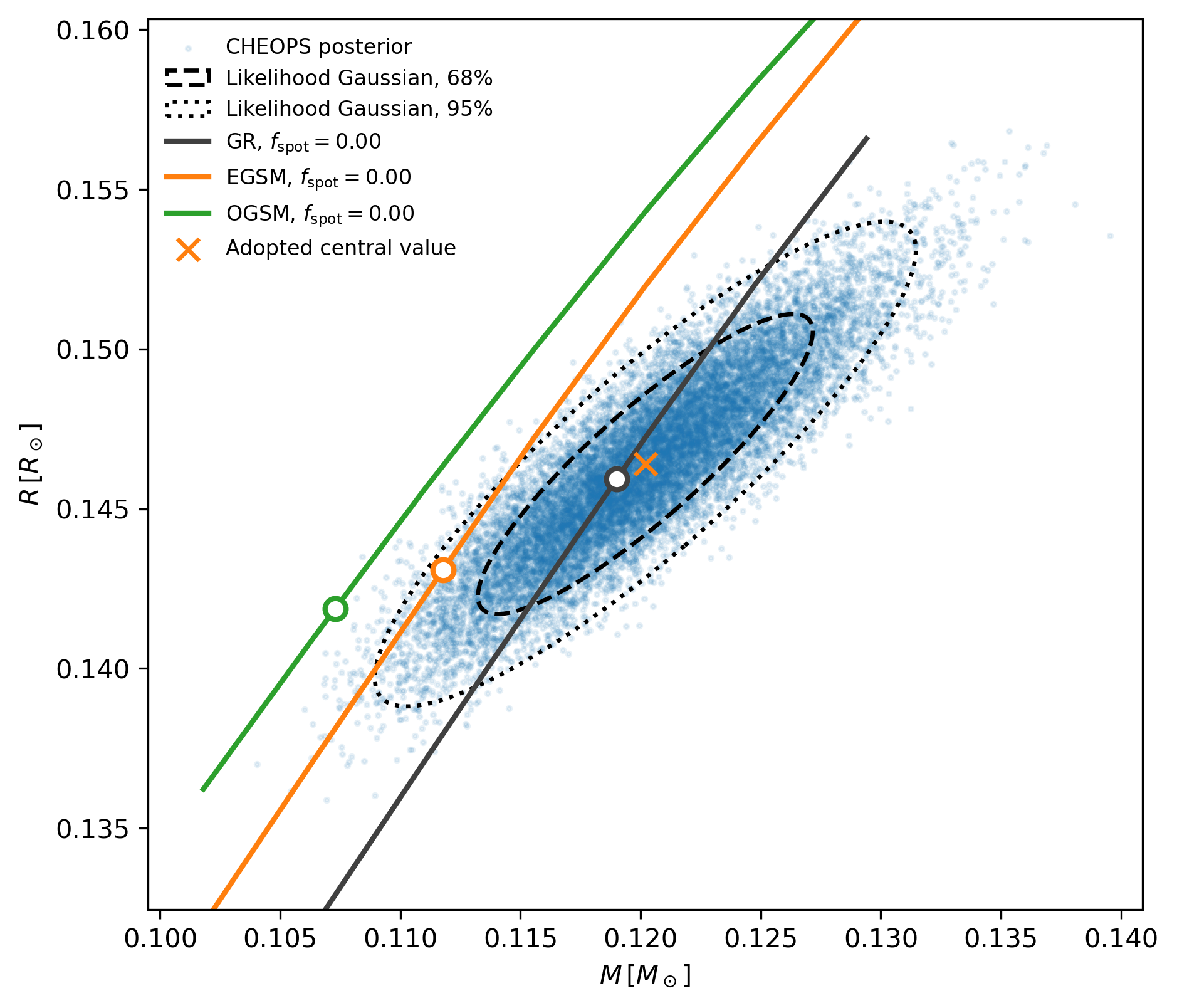}
    \caption{
    Mass--radius constraints for J2343+29.~The blue points show samples from the CHEOPS posterior, while the dashed and dotted contours show the $68\%$ and $95\%$ regions of the correlated Gaussian likelihood adopted in our analysis.~The solid curves show the theoretical mass--radius relations for constant $G=G_0$, EGSM, and OGSM at the independently profiled spot covering fraction; for J2343+29 all three solutions lie at the physical boundary $f_{\rm spot}=0$.
    The open circles mark the stellar masses that minimize the correlated mass--radius $\chi^2$ along each model relation, and the orange cross marks the adopted observational central value.~The raw posterior is well described by the Gaussian approximation.~The positive-$G$ benchmarks displace the model relation relative to the data in a direction with a substantial component across the narrow axis of the observational covariance, leading to an increasing likelihood penalty from constant $G$ to EGSM and OGSM.
    }
    \label{fig:j2343}
\end{figure}

The origin of this leverage is illustrated in Figure~\ref{fig:j2343}. The raw CHEOPS mass--radius posterior is well described by the correlated Gaussian likelihood adopted in our analysis, while the positive-$G$ benchmark model loci are displaced relative to the data in a direction with a substantial component across the narrow axis of the posterior. After profiling over the stellar mass, this produces $\chi^2=4.71$ and $12.55$ for J2343+29 at the EGSM and OGSM benchmarks, respectively, compared with $\chi^2=0.095$ for constant $G$.

As a further check, we tested the sensitivity of the bootstrap calibration to the nuisance-parameter generating point. For each system whose conditional benchmark fit lies at the physical boundary $f_{{\rm spot},i}=0$, we shifted the generating point to the largest available spot covering fraction satisfying $\Delta\chi_i^2\leq1$, re-profiling the stellar mass on the corresponding model curve; if no positive value satisfied this criterion, the boundary value was retained. This provides a one-sided stress test in which the boundary nuisance parameters are displaced into the interior while remaining compatible with the individual stellar data. For the EGSM benchmark, $11$ of $50{,}000$ realizations exceeded the observed likelihood-ratio statistic, compared with $47$ of $50{,}000$ for the nominal bootstrap. In a small number of realizations the profiled solution reaches the edge of the available stellar-mass grid; conservatively treating every case that could exceed the observed statistic after extending the mass support as a possible exceedance raises this count only to $15$. For the OGSM benchmark, none of the $50{,}000$ stress-test realizations produced a definite exceedance, with only one possible exceedance under the same mass-support criterion. Since $50{,}000$ realizations do not precisely resolve the OGSM tail, we use these calculations only as a sensitivity test. We find no evidence that this data-compatible displacement of the boundary nuisance parameters weakens either benchmark exclusion.

\section{Conclusions}
\label{sec:concs}

In this work, we studied the effects of a rapid transition in Newton's constant on the observable properties of red dwarf stars, motivated by proposals that invoke such a transition as a resolution of the Hubble tension. These objects have long thermal timescales, with radius-relaxation times of order $10^8$ yr, allowing an appreciable memory of a transition occurring 73--130 Myr ago to remain at the present day. Using both analytic and numerical models, we found that a transition from a larger (smaller) value of $G$ to its present-day value leaves red dwarfs inflated (deflated) relative to the corresponding constant-$G$ prediction. Comparing our predictions with eight red dwarfs in detached eclipsing binaries for which mass--radius covariance information is available, we find that constant $G$ provides the best fit for both transition times considered. The EGSM and OGSM benchmarks proposed to address the Hubble tension are excluded at significances of $3.1\sigma$ and $4.0\sigma$, respectively.~The exclusions are sensitive to J2343+29, particularly for OGSM:~omitting this system reduces the calibrated significances to $2.45\sigma$ and $1.90\sigma$ for EGSM and OGSM, respectively, while leaving the preferred value at $\Delta G/G_0=0$.

\section*{Acknowledgments}

HD is supported by Royal Society University Research Fellowship grant 211046.~Our simulations were run on the University of Hawai\okina i's high-performance supercomputer KOA.~The technical support and advanced computing resources from University of Hawai\okina i Information Technology Services – Cyberinfrastructure, funded in part by the National Science Foundation MRI award \#1920304, are gratefully acknowledged.~

\section*{Data Availability}

The data and software required to reproduce the results of this work are openly available in the Zenodo repository \href{https://zenodo.org/records/23093223}{https://zenodo.org/records/23093223}, Ref.~\cite{sakstein_2026_23093223}. The repository contains the \textsc{MESA} inputs used to generate the stellar models, reduced likelihood-ready model grids and observational inputs, the Python analysis and plotting code, and the bootstrap results used for the statistical calibration.

\appendix

\section{Atmospheric Boundary Condition for Fully Convective Stars}
\label{sec:appendix}

In this appendix we derive equation~\eqref{eq:Teff49}.~We accomplish this by introducing a polytropic model for the convective interior and matching this to the boundary condition at the photosphere.

\subsection{Convective Interior}

Since the interior is fully convective, we can describe it using an $n=3/2$ polytrope i.e.,
\begin{equation}
    P = K \rho^{5/3},
    \label{eq:polytropic_eos}
\end{equation}
where $K$ is the polytropic constant.~The Lane-Emden relations give
\begin{equation}
    K \propto G R M^{1/3}.
    \label{eq:K_scaling}
\end{equation}

Near the photosphere, the ideal-gas equation of state (see Eq.~\eqref{eq:ideal_gas_EOS}) gives
\begin{equation}
    \rho_{\rm ph}\propto \frac{P_{\rm ph}}{T_{\rm ph}}.
    \label{eq:ideal_gas_ph}
\end{equation}
Substituting Eq.~\eqref{eq:ideal_gas_ph} into
Eq.~\eqref{eq:polytropic_eos} and rearranging gives
\begin{equation}
    P_{\rm ph}
    \propto
    K^{-3/2}T_{\rm ph}^{5/2}.
    \label{eq:Pph_interior_K}
\end{equation}
Eliminating $K$ using Eq.~\eqref{eq:K_scaling}, the pressure at the photosphere is
\begin{equation}
    P_{\rm ph}^{\rm int}
    \propto G^{-3/2}R^{-3/2}M^{-1/2}T_{\rm eff}^{5/2},
    \label{eq:Pph_interior}
\end{equation}
where we set $T_{\rm ph}=T_{\rm eff}$.

\begin{table*}[t]
\centering
\caption{Observational properties of the red dwarf stars used in this work.~The mass--radius correlation coefficient $\rho_i$ is determined from the posterior samples of Ref.~\cite{2024MNRAS.528.5703S}.}
\label{tab:sample}
\begin{tabular}{lcccccccc}
\hline\hline
Target &
$M/\msun$ &
$\sigma_M/\msun$ &
$R/\rsun$ &
$\sigma_R/\rsun$ &
$\mathrm{[Fe/H]}$ &
$\sigma_{\mathrm{[Fe/H]}}$ &
$\rho_i$ &
Reference \\
\hline
J0540-17 & 0.1633 & 0.0058 & 0.1949 & 0.0032 & $-0.040$ & 0.050 & 0.716 & \cite{2024MNRAS.528.5703S} \\
J0719+25 & 0.1584 & 0.0055 & 0.1917 & 0.0029 & $\phantom{-}0.040$ & 0.050 & 0.548 & \cite{2024MNRAS.528.5703S} \\
J0941-31 & 0.2173 & 0.0078 & 0.2365 & 0.0036 & $\phantom{-}0.078$ & 0.069 & 0.696 & \cite{2024MNRAS.528.5703S} \\
J0955-39 & 0.2211 & 0.0080 & 0.2327 & 0.0030 & $-0.240$ & 0.080 & 0.568 & \cite{2024MNRAS.528.5703S} \\
J1928-38 & 0.2703 & 0.0091 & 0.2692 & 0.0057 & $-0.009$ & 0.042 & 0.859 & \cite{2024MNRAS.528.5703S} \\
J2046+06 & 0.1769 & 0.0062 & 0.2055 & 0.0025 & $\phantom{-}0.000$ & 0.048 & 0.898 & \cite{2024MNRAS.528.5703S} \\
J2315+23 & 0.2309 & 0.0099 & 0.2521 & 0.0034 & $\phantom{-}0.020$ & 0.050 & 0.534 & \cite{2024MNRAS.528.5703S} \\
J2343+29 & 0.1202 & 0.0046 & 0.1464 & 0.0031 & $\phantom{-}0.110$ & 0.050 & 0.883 & \cite{2024MNRAS.528.5703S} \\
\hline
\end{tabular}
\end{table*}

\subsection{Photospheric Boundary Condition}

Hydrostatic equilibrium in a thin atmosphere implies
\begin{equation}
    \frac{dP}{d\tau} = \frac{g}{\kappa},
\end{equation}
where $\tau$ is the optical depth and
\begin{equation}
    g=\frac{GM}{R^2}
\end{equation}
is the surface gravity.~The photosphere is defined by $\tau=2/3$, which gives the scaling relation
\begin{equation}
    P_{\rm ph}
    \sim
    \frac{g}{\kappa_{\rm ph}}.
    \label{eq:photospheric_hse}
\end{equation}
For the cool atmospheres of low-mass stars, the opacity is dominated by H$^{-}$ absorption, which can be approximated by \cite{2012sse..book.....K}
\begin{equation}
    \kappa \propto \rho^{1/2}T^9.
    \label{eq:opacity_rho_T}
\end{equation}
Substituting this into~\eqref{eq:photospheric_hse} one finds
\begin{equation}
    P_{\rm ph}
    \sim
G^{2/3}M^{2/3}R^{-4/3}T_{\rm eff}^{-17/3}.
    \label{eq:Pph_atmosphere}
\end{equation}

\subsection{Atmosphere--Interior Matching}

Imposing that the expressions for the pressure at the photosphere match between
Eqs.~\eqref{eq:Pph_atmosphere} and~\eqref{eq:Pph_interior} yields equation \eqref{eq:Teff49}.

\section{Observed Stellar Sample}
\label{app:data}

Table~\ref{tab:sample} lists the eight red dwarf stars used in this work.~The stellar masses, radii, metallicities, and associated uncertainties are taken from Ref.~\cite{2024MNRAS.528.5703S}.~The mass--radius correlation coefficients $\rho_i$ are calculated from the posterior samples provided by the authors, as described in Section~\ref{sec:method}.

\bibliography{refs}

\end{document}